# Goya's Artwork Imaging with Terahertz Waves

C. Seco-Martorell,[1,*] V. López-Domínguez,[1] G. Arauz-Garofalo,[1] A. Redo-Sanchez,[2]

J. Palacios, and J. Tejada[1]

[1]Departament de Física Fonamental, Facultat de Física, Universitat de Barcelona, c. Martí i Franquès 1, Planta 4, Edifici Nou, 08028 Barcelona, Spain

[2]Zomega Terahertz Corporation, East Greenbush, NY 12061, USA

## Abstract

In this paper we use a Terahertz (THz) time-domain system to image and analyze the structure of an artwork attributed to Spanish artist Goya painted in 1771. The THz images show features that cannot be seen in the optical or X-ray images that providence evidence of its authenticity, which is validated by other independent studies. For instance, a feature with a strong resemblance with one of Goya's known signatures is seen in the THz images that cannot be detected otherwise. This paper shows the feasibility of using THz imaging in the field of authentication and characterization of artworks. This may be of great importance in cases where both the signature and specific features that are critical for the authentication of an artwork remain hidden to visible, infrared, or X-ray inspection methods.

Keywords: terahertz, image reconstruction techniques, imaging,

## INTRODUCTION

Terahertz (THz) radiation covers the part of electromagnetic spectrum between the microwave and far infrared bands (0.1 to 10 THz). THz has been extensively studied as a tool for spectroscopy in security [1]-[4] and pharmaceutical applications [5]-[8]. THz waves can also be used as a non-destructive evaluation tool because they can penetrate many materials (see through), such as plastic, paper, cardboard, canvas, and textiles, and their short wavelength allow generating images with spatial resolutions in the order of mm to sub-mm. For instance, THz imaging has been applied to study defects in insulation materials, composites, and in medical studies for cancer diagnosis. These features (e.g. see-through and sub-mm resolution) make THz technology particularly interesting in applications for the inspection of cultural heritage. The use of THz radiation in the field of culture heritage inspection began in the late 1990 decade [9]. However, it has been during the past recent years that different types of studies have been reported, such as imaging of Egyptian papyrus, THz spectroscopy of ancient pigments, and THz imaging and layer analysis of paintings and archeological objects [10]-[14]. In this paper, we use a THz time-domain system to image and analyze the structure of the famous artwork "Sacrifice to Vesta" painted by Spanish artist Francisco de Goya y Lucientes (Fuendetodos 1746 - Bordeaux 1828) in 1771. In particular, this paper demonstrates the potential of THz imaging to verify the authenticity of artwork pieces through the detection of hidden features that are not detectable by other means such as optical, X-ray, or infrared. For the inspection, we use a commercial Mini-Z THz time-domain system manufactured by Zomega Terahertz Corporation that can be used for both imaging and spectroscopy applications.

**SAMPLE AND EXPERIMENTAL**

The painting, or sample, was inspected with a Mini-Z THz time-domain system configured for imaging in reflection geometry (Fig. 1a). The Mini-Z implements a pump-probe scheme to generate and detect the THz pulses. The pump beam generates the THz pulse and it is delayed respect the probe beam, which is used to detect the pulse. The Mini-Z includes a femtosecond laser that pumps a photoconductive antenna (PCA) to generate a THz pulse (pump beam). The THz pulse is focused onto the sample surface using a high-density polyethylene (HDPE) lens with 1" diameter and 1" focal length. The reflected beam is collected by the same lens and separated from the incoming beam by a beam splitter. The detection of the reflected beam is realized by electro-optical sampling with the probe beam. The shape and position of the pulses reflected from the sample are recorded in the time-domain in the so-called waveform. This waveform provides information about the layered structure of a sample similar to the information provided by an ultrasound system.

Because the Mini-Z focuses the THz pulse on a single spot, the sample is set on a XY motorized stage so that the image can be generated in a raster scan mode. A manual micrometer controls the distance of the painting respect the Mini-Z in the perpendicular direction of the incident beam (Z-axis). A computer controls the stages and records the waveform at each position of the canvas, generating a cube of data. The scan duration depends on the desired resolution and the dimensions of the area being imaged and it could vary from a few hours to one week with waveform acquisition rates around 2 Hz. In this paper, the images are taken with a step size of 1 mm. Due to the dimensions of the painting and limited range of the XY stage, the painting was imaged in two half sections of 165 mm x 240 mm each. We measured one half first and then we rotated the painting to measure the other half. Furthermore, each section was imaged in eight different subsections of 82.5 mm x 60 mm so that the computer could handle the size of the data file being generated. The images from each subsection were stitched together to create the image of the entire painting. In the THz images, the two main sections can be identified through different contrast due to slight changes in the orientation of the paint that changed the maximum reflected power.

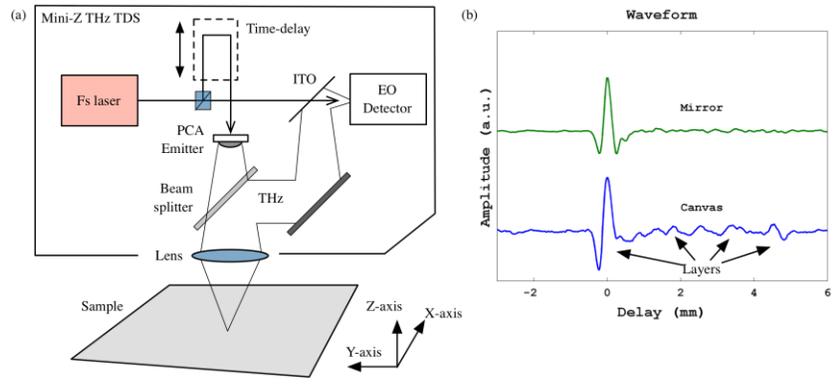

Fig. 1. Experimental layout.

Two analysis modes are possible with data provided by a THz time-domain system: structural and chemical. In the chemical analysis mode, the frequency components of the waveform (frequency-domain) can be analyzed via a Fourier transform. The spectrum of the waveform can provide information about specific resonances, or fingerprints, associated to particular pigments. In the structural mode, time of flight data is used to reconstruct the layered structure of the canvas and reveal features that are not visible because they are present at deeper layers of the sample. Fig. 1b shows a waveform corresponding to a point in the canvas in which different layers are seen as replicas of the main peak. A waveform corresponding to the reflection of a perfect single surface (mirror) that does not contain such replicas is also shown for comparison. In this paper, we focus on the structural analysis by analyzing the amplitude of the THz pulse across the painting area. An analysis in the frequency domain is also performed to check the relationship between detected features and the resolution at different wavelengths.

The title of the inspected artwork is "Sacrifice to Vesta" (Fig. 2a). This painting is an oil on a canvas with dimensions of 33 cm by 24 cm that has been published as a Francisco de Goya's work in almost hundred art history publications and displayed in numerous exhibitions. Currently, the paint belongs to a private collection. This important work was painted by the young artist in 1771, almost certainly during his short but seminal visit to Rome. The significance of "Sacrifice to Vesta" lies not only in its aesthetic beauty and technical accomplishment, but also in its status as the earliest known signed and dated work by Goya. For this reason, this painting plays a key role in the understanding of the artist's early production and it provides an invaluable insight into the least documented phase of his long and brilliant career. In July 2007, a detailed investigation undertaken by leading Goya experts [15] endorsed the attribution to Goya.

Goya's interest in the antique while in Rome is attested by his drawings of classical sculptures in his "Cuaderno Italiano". The cult of Vesta, the Goddess of fire and the domestic hearth, was widely celebrated in Rome. While "Sacrifice to Vesta" still reveals Goya at a formative stage in his career, it illustrates the speed of his artistic development during his stay in Rome in 1771. A slight naivety can still be detected in isolated parts of the composition such as the drawing of the two female figures behind the altar. However the brushwork is remarkably vigorous and confident, as seen in the spontaneous application of highlights to the drapery of the figure of the vestal virgin on the left of the scene and in the swirling clouds and moving trees. The figures themselves are depicted with a remarkable sense of monumentally and volume, achieved through the artist's highly skilled treatment of light which at once defines texture and form. The painting's boldness and spontaneity hint at the beginnings of a genius and hold out the promise of great things to come. In its overall directness and immediacy the work ranks among Goya's finest and most beautiful early creations.

**RESULTS**

Fig. 2b and 2c show different superposition between optical and THz amplitude images varying the degree of transparency. In the THz images we have plotted the maximum amplitude value of the waveform for each pixel. Other possible images can be obtained by plotting the position of the peak of the waveform or generating sections at different depths of the sample. However, amplitude images already display a great deal of structure of the painting that is not obvious by optical inspection. The THz images exhibit a strong correlation with the visible image. Some elements and figures of the painting, such as the priest and the arm and the head of Vesta (the woman on the left) can be clearly recognized in THz images.

Fig. 2b shows a superposition of 50% visible 50% THz, and Fig. 2c corresponds to a 100% THz image with no visible image superposition. The difference in reflectivity in areas of the painting is associated to the reflectivity of the pigments. This different reflectivity could be related with the metal content of the pigment. For instance, pigments with metallic content would show a stronger THz reflectivity than pigments with no metallic content.

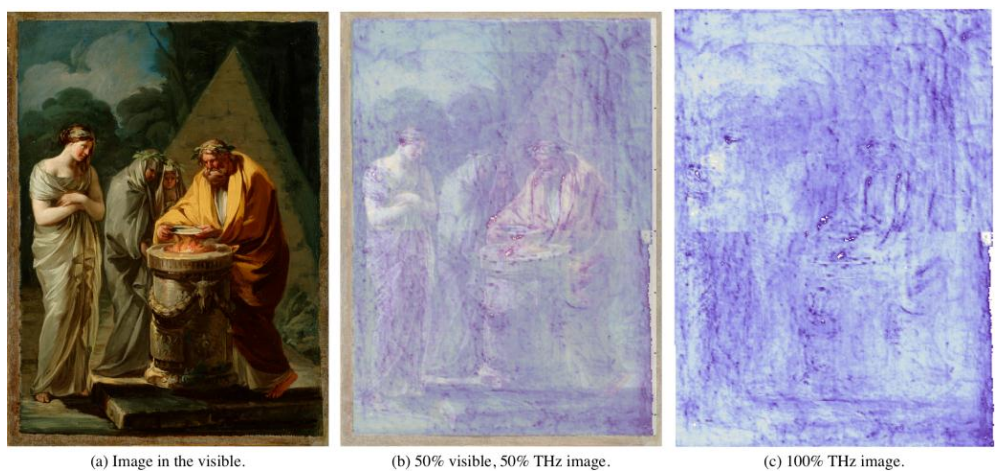

(a) Image in the visible.    (b) 50% visible, 50% THz image.    (c) 100% THz image.

Fig. 2. Composition of "Sacrifice to Vesta" at different transparency levels of visible and THz images.

The study of the THz images provides an insight on the painting technique and the detection of features that cannot be seen in the visible image. In particular, the THz image provides a "texture feel" of the painting in which the intensity of the stroke, the density of the pigments, and structural features of the canvas all come together (Fig. 3). For example, in the left upper corner in the THz image, a horizontal line and an arc are clearly seen. The horizontal line is attributed to the interaction of the wooden frame with the canvas, and the arc corresponds to some sort of mechanical defect of the canvas according to the evaluation of art experts. In the upper right corner, we can see wrinkles that could be caused by several effects such as mechanical tensions on the canvas (shrinkage or stretching), deterioration of pigments due to heat, erosion of the edges of the canvas due to fungi or the action of insects. In the THz image we can also distinguish the main strokes that give shape to the figures in the artwork.

However, the most surprising feature is found at the bottom right of the painting, which could correspond to Goya's signature. Goya signed his paintings with different signatures [16] and, one of them (Fig. 3b) could correspond to the observed traces. Enhancing the image on that particular area (Fig. 3c), we can clearly distinguish the "ya" and a feature in front of them that could correspond to the "G". However, the "o" cannot be detected. Fig. 3d and Fig. 3e show the images resulting from computing the Fourier transform on the waveform and, therefore, provide the amplitude distribution at different frequency components. As expected, the alleged signature can be seen at high frequencies such as 0.8 THz (Fig. 3d) but it cannot be observed at the lower frequency of 0.46 THz (Fig. 3e) because lower frequencies are not able to resolve the feature due to longer wavelengths.

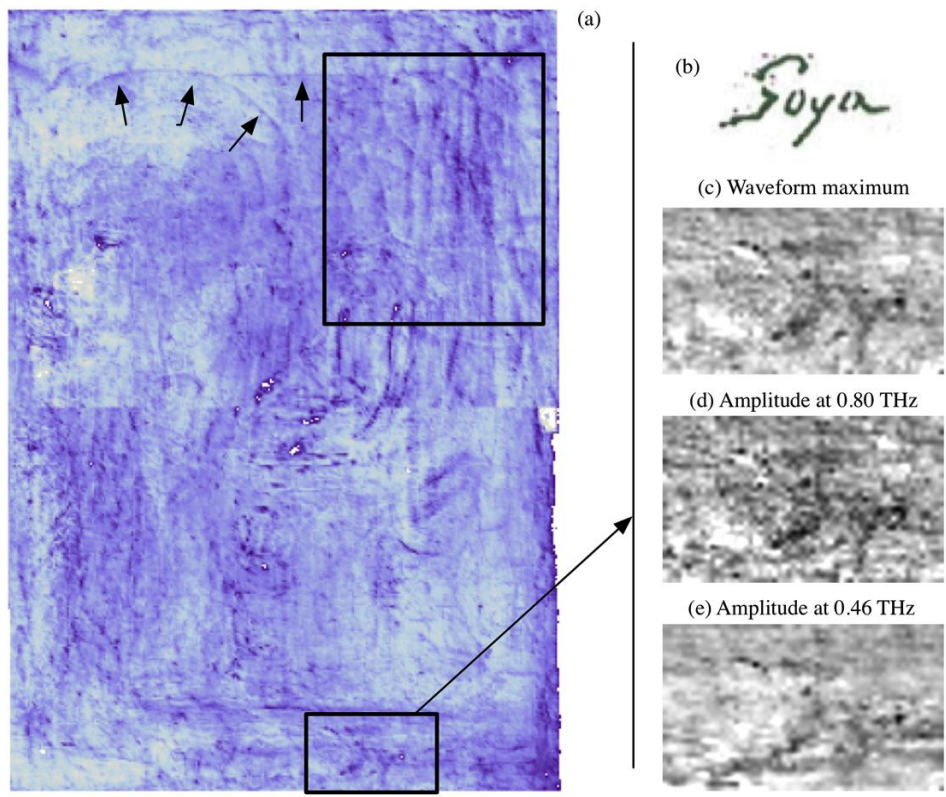

Fig. 3. (a) Structural features in the THz image and signature area. (b) Known signature of the artist. (c) Alleged signature resulting from finding the maximum amplitude in the time-domain data. (d) Image of the alleged signature in the frequency domain at 0.8 THz. (e) Image

of the same area in the frequency domain at 0.46 THz in which the alleged signature is not visible due to larger wavelength than that corresponding to 0.8 THz.

**DISCUSSION AND CONCLUSION**

The painting has been subjected to X-Ray (Fig. 4) and infrared studies [17]. Analyzing X-ray image in Fig. 4, the authors of the study conclude that the canvas is not damaged because no gashes neither tears are observed. On the other hand, the wrinkles seen in the THz image on the upper right corner may indicate a slight deterioration of the painting that, although it does not compromise the integrity of the painting at present time, it may require some sort of restoration operation in the future. The authors studying the X-ray also claim that the X-Ray image shows modifications that the artist had done during the execution of the paint respect the final composition. They emphasize that vertical strip at the left of the female figure may indicate a correction in the position of the body, the face of the woman located at the center was initially conceived in three quarters while the final composition is nearly in profile. However, the X-Ray image does not show any evidence of the alleged signature that is seen in the THz images.

It is likely that the signature was written using a pencil (basically carbon) and that the painting was covered by a top layer of finishing varnish that turned dark over time, hiding the signature to optical inspection. The fact that the signature cannot be seen in the X-ray image could be explained considering that the atomic weight of the carbon (signature) and surrounding canvas and paint is very similar and, thus, X-ray is mostly insensitive to this variation. On the other hand, THz waves are more sensitive to molecular composition and different reflectivity of carbon and the surrounding canvas can provide the mechanism for the detection of the signature in the THz image.

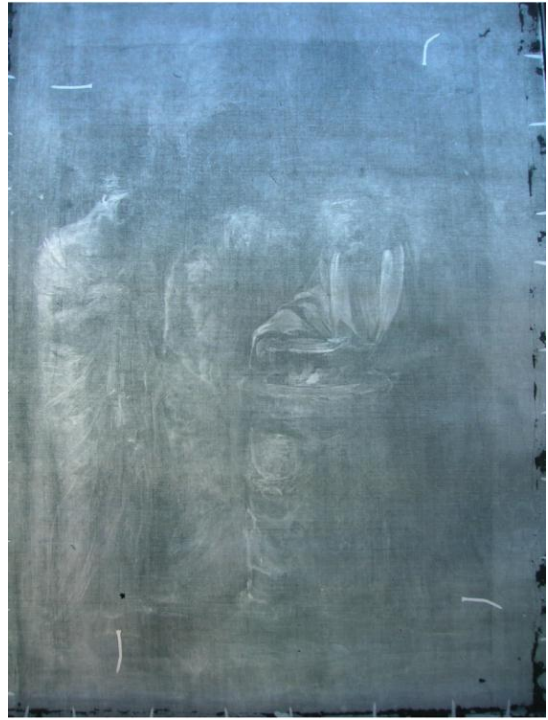

Fig. 4. X-ray image in which alleged signature is not visible nor wrinkles and defects on the upper third sector.

Therefore, we can conclude that the THz images provide complementary information to X-ray images and optical inspection because it can unveil features (hidden sketches and signatures, wrinkles, stroke style, etc.) that cannot be evaluated otherwise. Such features can support evaluation studies of paintings in relation to authorship authentication, structural integrity assessment and restoration needs. Besides the use of THz technology to analyze the structural properties of the paintings, THz technology has also the potential to study the chemical composition of the painting by analyzing the information provided by the waveform in the frequency-domain. This type of spectroscopy report requires the creation of a pigment spectroscopy database and proper calibration protocol that are outside of the scope of this paper.

**ACKNOWLEDGMENTS**

C. S.-M. would like to thank A. Pérez-Obiol for his technical support, and Fundación Goya en Aragón and Grup de Magnetisme (Universitat de Barcelona) for the financial support. V.L.-D. is grateful to Fundación Goya en Aragón and Grup de Magnetisme (Universitat de Barcelona) for the financial support. J. P. wishes to thank the assistance provided by Grup de Magnetisme (Universitat de Barcelona). G. A.-G. and J. T.-P. appreciate financial support from ICREA Acadèmia, Fundación Goya en Aragón and Universitat de Barcelona.

*Electronic address: cseco@ubxlab.com

[1] K. Kawase, H. Hoshina, A. Iwasaki, Y. Sasaki, T. Shibuy, "Mail Screening Applications of Terahertz Radiation," Electron. Lett. **46**, S66–S68 (2010).

[2] J. Chen, Y. Chen, H. Zhao, G. J. Bastiaans, X.-C. Zhang, "Absorption Coefficients of Selected Explosives and Related Compounds in the Range of 0.1-2.8 THz," Opt. Express **15**, 12060–12067 (2007).

[3] Y. Shen, T. Lo, P. F. Taday, B.E. Cole, W. R. Tribe, and M. C. Kemp, "Detection and Identification of Explosives Using Terahertz Pulsed Spectroscopic Imaging," Appl. Phys. Lett. **86**, 241116 (2005).

[4] K. Yamamoto, M. Yamaguchi, F. Miyamaru, M. Tani, M. Hangyo, T. Ikeda, A. Matsushita, K. Koide, M. Tatsuno, and Y. Minami, "Noninvasive Inspection of C-4 Explosive in Mails by Terahertz Time-Domain Spectroscopy," Jpn. J. Appl. Phys. **43**, L414–L417 (2004).

[5] A. J. Fitzgerald, B. E. Cole, and P. F. Taday, "Nondestructive Analysis of Tablet Coating Thicknesses Using Terahertz Pulsed Imaging," J. Pharm. Sci **94**, 177–183 (2005).

[6] C. J. Strachan, P. F. Taday, D. A. Newnham, K. C. Gordon, J. A. Zeitler, M. Pepper, and T. Rades, "Using Terahertz Pulsed Spectroscopy to Quantify Pharmaceutical Polymorphism and Crystallinity," J. Pharm. Sci **94**, 837–846(2005).

[7] J. A. Zeitler, P. F. Taday, D. A. Newnham, M. Pepper, K. C. Gordon, and T. Rades, "Terahertz Pulsed Spectroscopy and Imaging in the Pharmaceutical Setting - a Review," J. Pharm. Pharmacol. **59** 209–223 (2007).

[8] C. Gendre, M. Genty, M. Boiret, M. Julien, L. Meunier, O. Lecoq, M. Baron, P. Chaminade, and J. M. Pean, "Development of a Process Analytical Technology (PAT) for in-Line Monitoring of Film Thickness and Mass of Coating Materials During a Pan Coating Operation," Eur. J. Pharm. Sci. **43**, 244–250 (2011).

[9] J. B. Jackson, J. Bowen, G. Walker, J. Lebaune, G. Mourou, M. Menu, and K. Fukunaga, "A Survey of Terahertz Applications in Cultural Heritage Conservation Science," IEEE Trans. Terahertz Sci. **1**, 220–231 (2011).

[10] J. Labaune, J. B. Jackson, S. Pages-Camagna, I. N. Duling, M. Menu, and G. A. Mourou. "Papyrus Imaging with Terahertz Time Domain Spectroscopy," Appl. Phys. A **100**, 607–612 (2010).

[11] E. Abraham, A. Younus, J. C. Delagnes, and P. Mounaix, "Non-Invasive Investigation of Art Paintings by Terahertz Imaging," Appl. Phys. A **100**, 585–590 (2010).

[12] G. P. Gallerano, A. Doria, E. Giovenale, G. Messina, A. Petralia, I. Spassovsky, K. Fukunaga, and I. Hosako, "THz-ARTE: Non-Invasive Terahertz Diagnostics for Art Conservation," in *Proceedings of IEEE of Infrared, Millimeter and Terahertz Waves*, (Pasadenca, CA, 1998), pp. 1-2.

[13] E. Abraham, A. Younus, A. El Fatimy, J. C. Delagnes, E. Nguema, and P. Mounaix, "Broadband Terahertz Imaging of Documents Written with Lead Pencils," Opt. Commun. **282**, 3104–3107 (2009).

[14] K. Fukunaga, and M. Picollo, "Terahertz Spectroscopy Applied to the Analysis of Artists' Materials," Appl. Phys. A **100**, 591–597 (2010).

[15] Dr. M. Mena-Márquez and J. Wilson (private communication).

[16] A. Canellas-López, "Las firmas y rúbricas del Pintor Goya" (Real Academia de Nobles y Bellas Artes de San Luís, Zaragoza, 1991).

[17] M. Riera and N. Hernández (private communication, 2007).